\documentclass [twocolumn, aps, showpacs] {revtex4}
\usepackage{graphicx}
\usepackage{amsmath}
\usepackage{amssymb}
\usepackage{epsf}
\newlength{\upit}\upit=0.1truein

\newcommand{\ltappr}{{{\lower4pt\hbox{$<$} } \atop \widetilde{ \ \ \ }}}
\newlength{\bxwidth}\bxwidth=1.5 truein

\def\be{\begin{equation}}
\def\ee{\end{equation}}
\def\vs{\vspace}

\def\dg{^{\dagger}}

\newlength{\figwidth}
\newlength{\shift}
\newcommand \bea {\begin{eqnarray} }
\newcommand \eea {\end{eqnarray}}

\begin{document}
\draft
\title {Spin dynamics from Majorana fermions}

\author {W. Mao$^{1}$, P. Coleman$^{2}$, C. Hooley$^{3}$ and D. Langreth$^{2}$}
\affiliation{$^{1}$ Department of Physics and Astronomy, University of Stony
Brook, SUNY, Stony Brook, NY 11794-3800, U.S.A.}
\affiliation{$^{2}$ Center for Materials Theory, Rutgers University, Piscataway, NJ 08854-8019, U.S.A.}
\affiliation{$^{3}$ School of Physics and Astronomy, Birmingham University, Edgbaston, Birmingham B15 2TT, U.K.}
\date{5th May, 2003}
\begin{abstract}
Using the Majorana fermion representation
of spin-$1/2$ local moments, we show 
how it is possible to directly read off the dynamic spin
correlation and susceptibility from the one-particle
propagator of the Majorana fermion.  We illustrate our method by applying
it to the spin dynamics of  
a non-equilibrium quantum dot, computing the voltage-dependent spin
relaxation rate and showing 
that, at weak coupling, the
fluctuation-dissipation relation for the spin of a quantum dot is
voltage-dependent.  We confirm the 
voltage-dependent Curie susceptibility recently found
by Parcollet and Hooley [Phys.~Rev.~B {\bf 66}, 085315 (2002)].
\end{abstract}
\maketitle

The mathematical difficulties of representing spins in 
many body physics have long been recognized. 
The essence of the problem is that spin operators are non-abelian:
they do not obey
Wick's theorem and an expectation value of the product of many spin
operators cannot be decomposed into products of two-operator
expectation values, even within a free theory.

A conventional response to this difficulty is to
represent spins as
bilinears of 
fermions \cite{Abrikosov} or as bosons \cite{Schwinger}. One of the disadvantages of these approaches is that the Hilbert space 
of the fermions or bosons needs to be restricted by the application of
constraints \cite{constraints,c2,c3}.  Another difficulty  is the ``vertex problem'', which
arises in the context of spin dynamics and spin relaxation. 
Once the spins are represented as bilinears, the spin-spin correlation functions 
are represented by two-particle 
Green's functions.  The calculation of these quantities requires
a knowledge of both the four-leg vertex and the single-particle
Green's function. Typically, the vertex is simply neglected, or treated in 
a very  approximate fashion. 

An alternative approach is to take advantage of the anticommuting properties
of Pauli matrices, writing the spin operator
in terms of Majorana fermions \cite{tsvelik,history5,history6,majo,shastry}, 
\begin{equation}\label{majrep}
 \vec S = - \frac{i}{2} \vec \eta \times \vec \eta,
\end{equation}
where $\vec \eta = (\eta^1,\eta^2,\eta^3)$ is a triplet of Majorana fermions
which satisfy $\{\eta^a, \eta^b\}=  \delta^{ab}$. 
This representation does not require the imposition
of a constraint:\ the fact that ${\vec S}^2 = 3/4$ follows directly from the
operator properties of the Majorana fermions.
In this letter, we show how this 
representation
also solves the vertex problem. To demonstrate this, we 
employ an alternative
derivation \cite{miranda} of the Majorana spin representation.
Consider a spin-1/2 operator $\vec{S}$ with 
dynamics described  by a Hamiltonian $H$.  Let us 
now introduce a single Majorana fermion $\Phi $ which lives in a
completely different Hilbert space, commuting with $\vec S$
and $H$. 
It follows that $\Phi$ is 
a fermionic constant of motion, 
$d\Phi/dt = -i [ H,\Phi] = 0$: an object of fixed magnitude
$\Phi^2=1/2$  which anticommutes with
all other fermion operators. We may now identify
$\vec{\eta}$ in (\ref{majrep})
with the operator identity
\begin{equation}\label{main}
\vec \eta = 2 \Phi \vec S.
\end{equation}
We may confirm that  $\{\eta^a, \eta^b\}= 
\delta^{ab}$ using the 
anticommuting algebra of spin-1/2 operators $\{S^a,S^b\} =
\frac{1}{2}\delta^{ab}$. 
Furthermore, using the SU(2) algebra of spins,
$\vec\eta \times \vec \eta = 2 \vec S\times \vec S= 2i \vec S$, from which \eqref{majrep}
follows immediately. 
As a last step in the derivation, we note that the independent Majorana operator
can also be written in the form
$ \Phi \equiv -2i
\eta^1 \eta^2 \eta^3$, an object that can be verified to commute with expression \eqref{majrep}. 
(Notice, incidentally, that although it is true that $\vec{S} =
\Phi \vec{\eta }$, this expression is of limited use because
$\vec{\eta}$ and $\Phi$ are not independent fermions:\ they
commute, rather than anticommuting.)

The important, yet previously unemphasized 
feature brought out by this derivation is that 
the Majorana fermions and the  spin operator are proportional
to one another,  $\vec \eta \propto \vec S$, where the constant of
proportionality is a constant fermion.
From this fact, it follows that 
\begin{eqnarray}\label{duality}
\frac{1}{2}\eta^a(t_1) \eta^b(t_2) & = & 2
\Phi(t_1)\Phi(t_2) S^a(t_1)S^b(t_2) \cr
&=&   S^a(t_1) S^b(t_2).
\end{eqnarray}
This operator identity enables us to 
connect the spin correlation function to a {\it one}-particle
Majorana Green's function. 
Inserting 
commutators or anticommutators into (\ref{duality} ) and taking the expectation value,  we find
\begin{center}
\begin{tabular}{c c c}
correlation function & & response function \\
of spins & & of Majoranas \\
$\overbrace{-i \langle \{ S^a(t_1),S^b(t_2) \} \rangle}$
& = &
$\overbrace{-\frac{i}{2} \langle \{ \eta^a(t_1),\eta^b(t_2) \} \rangle}$, \\
\ & \ & \ \\
response function & & correlation function \\
of spins & & of Majoranas\\
 $\overbrace{-i\langle [ S^a(t_1),S^b(t_2) ] \rangle}$
& = &
 $\overbrace{-\frac{i}{2} \langle [ \eta^a(t_1),\eta^b(t_2) ] \rangle}$.
\\
\end{tabular}
\end{center}
The expectation of a spin anticommutator is a correlation
function, but its fermionic counterpart represents 
a response function.
Likewise, the expectation value of a spin commutator represents
a spin response function, but this is equal to a fermion
correlation function or ``Keldysh'' Green's function \cite{Keldysh}.
Thus the
correlation function of the Majorana fermions determines the
{\it response} function of the physical spins, and vice versa.

We may formalize this relationship,  writing
\begin{eqnarray}\label{links}
\underline{\chi }''(t_{1},t_{2})&=&\frac{i}{4}\underline{G}_{K} (t_{1},t_{2})\\
\underline{C } (t_{1},t_{2})&=&\frac{i}{4}\biggl[\underline{G}_{R}(t_{1},t_{2})
-\underline{G}_{A} (t_{1},t_{2})
\biggr]
\end{eqnarray}
where 
\begin{eqnarray}\label{}
(\underline{\chi }'')^{ab} (t_{1},t_{2})&=&
\frac{1}{2}\langle
[S^{a} (t_{1}),S^{b} (t_{2})]\rangle \\
(\underline{C})^{ab} (t_{1},t_{2})
&=& \frac{1}{2}\langle
\{S^{a} (t_{1}),S^{b} (t_{2})\}\rangle 
\end{eqnarray}
are the spin response and correlation functions and 
\begin{eqnarray}\label{define}
{G}_{K}^{ab} (t_{1},t_{2})&=& -i \langle [\eta ^{a} (t_{1}),\eta
^{b} (t_{2})]\rangle \\
{G}_{R}^{ab} (t_{1},t_{2})&=& - i \langle \{ \eta ^{a} (t_{1}),\eta
^{b} (t_{2})\} \rangle  \theta (t_{1}-t_{2}) \\
{G}_{A}^{ab} (t_{1},t_{2})&=& i \langle \{ \eta ^{a} (t_{1}),\eta
^{b} (t_{2})\} \rangle  \theta (t_{2}-t_{1})
\end{eqnarray}
are the Keldysh, retarded and  advanced Green's functions of the
Majorana fermion. 
For most purposes, we are interested in systems that are in thermal
equilibrium, or that have reached a non-equilibrium steady state, for which
the correlation and Green's functions are functions only of the time difference 
$t_1-t_2$.  In this case, we may transform (\ref{links}) into 
frequency space, writing
\begin{eqnarray}\label{link2}
\underline{\chi }'' (\omega)&=&\frac{i}{4}\underline{G}_{K} (\omega),\\
\underline{C } (\omega)&=&\frac{i}{4}\biggl[\underline{G}_{R}(\omega)
-\underline{G}_{A} (\omega)
\biggr].
\end{eqnarray}
Here, $\chi'' (\omega)={\rm Im} [\chi_{R}(\omega)]=  {\rm Im} [\chi (\omega+ i
\delta )]$ is the imaginary part of the retarded spin susceptibility. 

It is particularly useful to combine the
Majorana fermions into a conventional Dirac fermion, writing $f\dg \equiv \frac{1}{\sqrt{2}}(\eta^1 + i \eta^2)$, for which $\{f,f\dg\}=
1$. 
The $f$-fermion
is directly proportional to the spin raising operator
$f\dg = \sqrt{2}\Phi S^+$, so that 
$S^-(t)S^+(0)= f(t)f\dg(0) $. 
Recasting the steady state version of 
(\ref{links}) in terms of the raising and lowering
operators, and Fourier transforming the resulting expressions, we obtain
\begin{eqnarray}\label{identity}
C^{-+}(\omega) &=& \frac{\pi}{2} 
 A(\omega),\\
(\chi'')^{-+}(\omega) &=&\frac{\pi}{2} 
A(\omega) h(\omega).
\label{freq}
\end{eqnarray}
where $A = \frac{i}{2\pi} ( G_{R} -G_{A})=\frac{1}{\pi} {\rm Im}\,G_{A} (\omega)$
is the $f$ spectral function 
and 
\begin{equation}
h (\omega) \equiv \frac{G_{K}(\omega)}{G_{R}(\omega)-G_{A}(\omega)}.
\end{equation}
In equilibrium, the function $h (\omega)=h_{E}(\omega) \equiv \tanh
(\omega/2T)$ is determined by the fermionic fluctuation-dissipation
theorem. We recover the conventional bosonic fluctuation-dissipation
theorem as the inverse of $h_{E} (\omega)$:
\[
\frac{C^{-+} (\omega) }{(\chi'')^{-+} (\omega)} = \frac{1}{h_{E} (\omega)}=\coth
\left[\frac{\omega}{2T} \right].
\]
In non-equilibrium steady state conditions, $h (\omega)$ must be 
computed from first principles, as  a non-equilibrium fluctuation-dissipation theorem, but the inverse relation between the
spin and fermionic fluctuation-dissipation functions is preserved. 

We can apply the Kramers-Kronig relation to
determine the full dynamic susceptibility from (\ref{identity}), as
\begin{eqnarray}\label{susc}
\chi^{-+} (\omega)&=&  \int d\nu\frac{A (\nu
)h (\nu)}{\nu-\omega - i \delta },
\end{eqnarray}so that the static transverse susceptibility is given by
\begin{equation}\label{static}
\chi_{\perp } = \frac{1}{2}\chi^{-+} (0)=\int d\nu\frac{A (\nu
)h (\nu)}{2\nu }.
\end{equation}
Since 
$(\chi'')^{-+}(\omega)$ is the Fourier transform of the response
function $\frac{1}{2}\langle [S^-(t),S^+(0)]\rangle$, using 
$[S^-,S^+]= - 2 S^z$, we can evaluate the $z$-component of the magnetization
from 
\begin{eqnarray}\label{magnet}
\langle S^z \rangle = -\frac{1}{2}\chi^{-+} (t=0)=
- \frac{1}{2} \int d \omega A(\omega) h(\omega).
\end{eqnarray}
Thus from the fermion propagator one can read off both the spin
dynamics and the static  magnetization.

To illustrate this method in its simplest form, consider a spin-$1/2$ in
a magnetic field in the negative $z$ direction, for which
$H = B S^z$.  Written in terms of fermions, 
\begin{equation}
H =   - i \eta^1 \eta^2 B = B\left(f\dg f - \frac{1}{2}\right).
\end{equation}
The retarded $f$-Green's function is given by
\begin{equation}
G_{R}(\omega) = \frac {1}{\omega - B + i \delta},
\end{equation}
so that $A(\omega) =\delta(\omega- B)$, and 
from (\ref{identity}) it follows that 
\begin{eqnarray}\label{}
(\chi'')^{-+}(\omega) &=&  \pi \delta(\omega- B)\tanh
\left(\frac{B}{2T}\right),
\\
C^{-+}(\omega) &=& \pi \delta(\omega -B),
\end{eqnarray}
and from (\ref{magnet}), $\langle S^z \rangle = - \frac{1}{2}\tanh \frac{B}{2T}$,
recovering the Brillouin function. 

The utility of the method is its ability to handle 
both equilibrium and non-equilibrium situations. To illustrate this point,
consider a spin coupled to two conduction seas,
according to the Kondo Hamiltonian
\begin{eqnarray}\label{kondo}
H &=&  \sum_{k \lambda\sigma } 
\epsilon_{k\lambda}c \dg _{k\lambda\sigma }c  _{k\lambda\sigma}
+ \sum_{\lambda, \lambda'}H_{\lambda\lambda'}\\
H_{\lambda \lambda'}&=&
\sum_{k,k'}J_{\lambda \lambda'}c \dg _{k\lambda
\alpha }\vec{\sigma }_{\alpha \beta}c _{k'\lambda'\beta}\cdot \vec{S}
\end{eqnarray}
where the terms $H_{\lambda\lambda'}$ ($\lambda=L,R$) 
describe the electron ``co-tunneling'' between lead $\lambda $ and
$\lambda'$ that is mediated by spin exchange with the spin $\vec{S}$. 
This model has been used to describe the
low energy physics of a quantum dot. Even when perturbation methods are
applied to this model, it is difficult to directly extract the
spin dynamics. The Majorana method permits the spin dynamics to be
computed 
perturbatively
in the couplings $J_{\lambda \lambda'}$, without any approximation to the spin vertex,  even when the two leads
are at different voltages. 

The Green's functions in zero field are now given by 
\begin{eqnarray}\label{gfn}
G_{R,A}(\omega) &=& \frac{1}{\omega -  \Sigma_{R,A}(\omega)},\\
G_K(\omega)&=& G_R(\omega)\Sigma_K(\omega) G_A(\omega),\end{eqnarray}
where $\Sigma_{R,A,K}(\omega)$ are the retarded, advanced, and Keldysh self-energies of 
the $f$-fermion, so that
\begin{equation}\label{}
h(\omega) = \frac{\Sigma_K(\omega)}{\Sigma_R(\omega)- \Sigma_A(\omega)} 
\end{equation}
reflects a change in the fluctuation-dissipation theorem. 
The spin-spin correlation function is thus given by
\begin{equation}\label{corr}
C^{-+} (\omega) = \frac{1}{4i}\left[
\frac{1}{\omega - \Sigma_{A}
(\omega)} 
-
\frac{1}{\omega - \Sigma_{R}
(\omega)} 
\right].
\end{equation}
Writing $\Sigma_{R} (\omega)= \Sigma' (\omega)- i\Gamma
(\omega)$, then at low frequencies 
\begin{equation}\label{lorentz}
C (\omega )= \frac{Z}{2} \frac{\Gamma_{0}}{\omega^{2}+ \Gamma_{0}^{2}}
\end{equation}
is Lorentzian, where $Z= (1 - d \Sigma '/ d\omega)^{-1}\vert _{\omega =
0}$ and $\Gamma_{0}= Z \Gamma (0)$ is the spin relaxation rate.

\begin{figure}
\begin{center}
\leavevmode
\hbox{\epsfxsize=8cm \epsffile{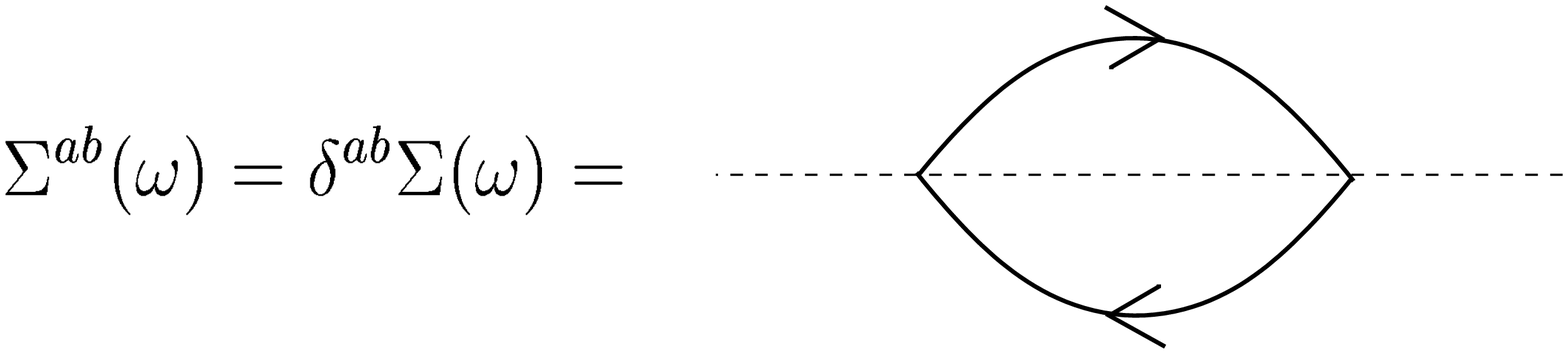}}
\end{center}
\caption{\small Leading order Feynman diagram for the Majorana fermion self-energy
$\Sigma_{ab} (\omega)$, 
which is isotropic
in spin indices in zero applied field. The
self-energy of the $f$-fermion $f\dg =\frac{1}{\sqrt{2}} (\eta
^{1}+i\eta ^{2})$ is given by $\Sigma (\omega)$.}
\label{fig1}
\end{figure}

To leading order in the coupling constant, 
we can compute the non-equilibrium self-energies by a simple
transformation  of the equilibrium self-energies. 
(The relevant Feynman diagram is shown in Fig.~\ref{fig1}.)
In equilibrium, the retarded self-energy, computed by an analytic
continuation of the imaginary time self-energy, is given 
by 
\begin{equation}\label{sigmar}
\Sigma_R(\omega) = 4\sum_{\lambda,\lambda'}(\bar J_{\lambda \lambda'})^2 \Upsilon(\omega),
\end{equation}
where $\bar J_{\lambda \lambda'}= \rho J_{\lambda \lambda'}$, and $\rho$
is the density of states per spin per lead.
\begin{equation}\label{phi}
\Upsilon(\omega) = \omega\left[
\psi\left(\frac{\omega}{2 \pi i T}\right) - \ln \left(\frac{D}{2 \pi T}\right)+ \frac{i \pi T}{\omega}
\right],
\end{equation}
where $\psi (z)= d \ln  \Gamma (z)/dz$ is the digamma function and $D$ is the bandwidth of each
lead, which has been introduced as a Gaussian cut-off in the Feynman
diagrams. 
The imaginary part of (\ref{phi}) is given by
\[
{\rm Im}\, \Upsilon (\omega)= -\frac{\pi \omega}{2}\coth \left[\frac{\omega}{2T} \right]
\]
so that the equilibrium Keldysh self-energy is given by 
\begin{equation}\label{}
\Sigma_{K} (\omega)= (\Sigma_{R}-\Sigma_{A})h_{E} (\omega)= -4\pi i\sum_{\lambda,\lambda'}({\bar J}_{\lambda\lambda'})^2 \omega.
\end{equation}
\begin{figure}
\begin{center}
\leavevmode
\hbox{\epsfxsize=8cm \epsffile{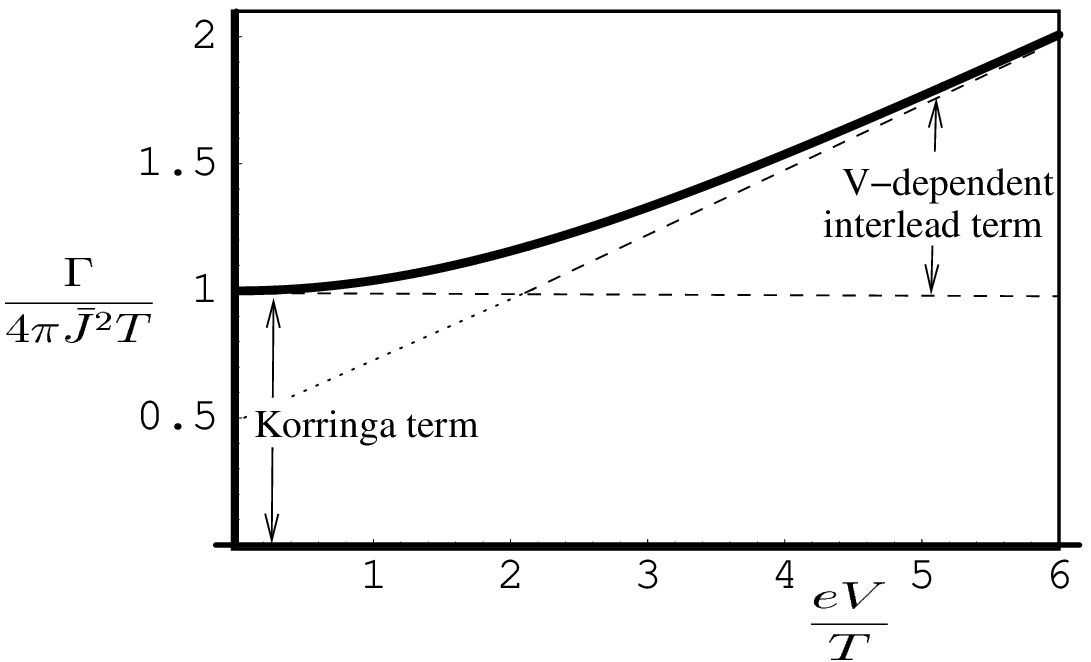}}
\end{center}

\vs{-5mm}
\caption{\small
Voltage-dependent
spin-relaxation rate, computed to leading quadratic order in the Kondo
coupling constant, showing the zero field, Korringa component, and the
the voltage-dependent inter-lead component. For 
the case chosen, $\bar J_{\lambda\lambda'}=\bar J$ ($\lambda, \lambda' \in \{R,L\}$).}
\label{fig3}
\end{figure}
The effect of applying a voltage to the leads can be incorporated into
the self-energies by noting that a voltage is equivalent 
to a gauge transformation  on the conduction electron fields, 
$c_{k\lambda\sigma }\rightarrow c_{k\lambda\sigma
}e^{i\mu_{\lambda}t}$, where $\mu_{\lambda}$ is the chemical potential
shift in the $\lambda=L,R$  lead. In a second-order calculation, this
gauge transformation can be incorporated by making the replacement
$\omega\rightarrow \omega - (\mu_{\lambda}-\mu_{\lambda'})$ in each
term of the self-energy, so that at finite voltage
\begin{eqnarray}\label{sigmane}
\Sigma_R(\omega) &=& 4\sum_{\lambda,\lambda'}(\bar J_{\lambda\lambda'})^2 \Upsilon[\omega-(\mu_{\lambda}-\mu_{\lambda'})],
\\
\Sigma_{K} (\omega)&=& -4\pi i \omega [ \bar J_{RR}^{2}+\bar
J_{LL}^{2}+2\bar J_{RL}^{2}],
\end{eqnarray}
where the voltage dependence cancels out of the second expression. 
Using (\ref{lorentz}) we may immediately read off the leading order
expression for the voltage-dependent spin relaxation rate of a quantum dot:
\begin{equation}\label{relaxation}
\Gamma_{0} (V,T) = 4 \pi\bar J_{RL}^{2}
\left[
2\alpha T +
eV \coth \left(\frac{eV}{2T} \right)
 \right],
\end{equation}
where we have introduced $eV = \mu_{L}-\mu_{R}$ and
$\alpha = (\bar J_{LL}^{2}+\bar J_{RR}^{2})/ 2\bar J_{RL}^{2}$. 
(See Fig.~\ref{fig3}.)
We recognize the $V=0$ limit
of (\ref{relaxation}) as the Korringa relaxation rate of a single
spin \cite{langreth72}. At finite $V$, the second term gives the
voltage-dependent spin relaxation
rate induced by the coupling between leads: this term is linear in temperature
for $eV \ll T$, but linear in voltage for $eV \gg T$. 
An identical result for ${\bar J}_{\lambda\lambda'} = {\bar J}$
was obtained in ref.~\cite{plihal01}.
We
can  also read off the voltage-dependent fluctuation-dissipation relation (see Fig.~\ref{fig2}),
\begin{equation}\label{}
h (\omega)= \frac{(1 + \alpha )\omega}{\frac{1}{2}[\phi
(\omega+eV)+\phi(\omega-eV)] + \alpha \phi (\omega)
},
\end{equation}
where $\phi (x)= {x}\coth ( x /2T)$,
so that by (\ref{static}) the static
susceptibility is given by
\begin{equation}
\chi (V,T)= \frac{1}{4T}\frac{(1+\alpha) }{[(eV/2T) \coth (eV/2T)+
\alpha ]},
\end{equation}
corresponding to a voltage-dependent Curie susceptibility. 
\begin{figure}
\begin{center}
\leavevmode
\hbox{\epsfxsize=8cm \epsffile{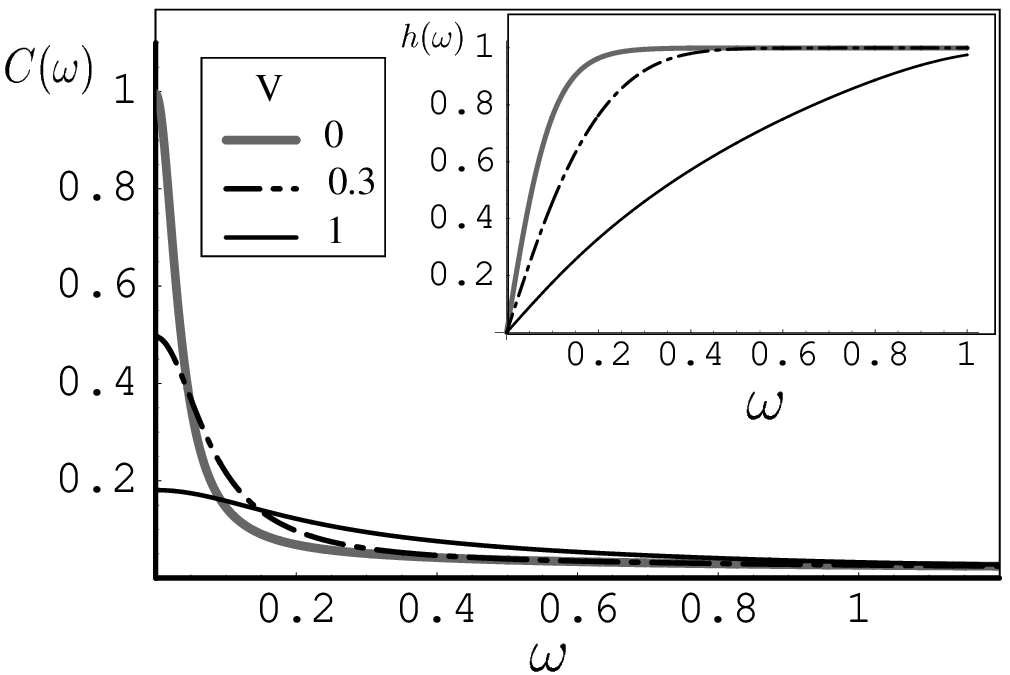}}
\end{center}

\vs{-5mm}
\caption{\small Voltage-dependent spin correlator $C(\omega)$ calculated for a
sequence of voltages $V=0,\,0.3,\,1$ at a temperature $T=0.05$. Here $\alpha =1$ and
$\bar J=0.2$.  Inset, fluctuation
dissipation function $h (\omega)$ for the same sequence of voltages.}
\label{fig2}
\end{figure}
This result
confirms earlier results of Parcollet and Hooley \cite{parcollethooley}, obtained by a direct
calculation of the magnetization using a self-consistent expression
for the Keldysh Green's function. 
Parcollet and Hooley also
obtain the finite field result $M (B)= \frac{1}{2}h (B,V)$,
which suggests that the voltage-dependent fluctuation-dissipation theorem is
independent of field at weak coupling.

Clearly, although this is beyond the scope of the current paper, 
the approach taken above can be extended to higher orders. An
interesting question that this may help answer
is whether the coherence of the Kondo effect is preserved at high
voltage bias for the (physical) antiferromagnetically coupled quantum
dot \cite{Kaminski:2000,Wen:1998,ColemanPH,Roschgroup}.

One of the enticing possibilities that this method offers is that
of extension to more complex, multi-impurity or even
lattice spin problems. 
The proportionality between spin and Majorana fermions can be extended
to these cases, merely by introducing
an independent Majorana fermion $\Phi _{j} $ for each spin site, and writing
$\vec{ \eta }_{j}= 2 \Phi _{j}\vec{S}_{j}$. The generalization of (3)
to a lattice is then
\begin{equation}\label{dualitylat}
S_{i}^a(t_1)S_{j}^b(t_2) = \frac{1}{2Z_{ij}}\eta^a_{i}(t_1) \eta^b_{j}(t_2)
\end{equation}
where $Z_{ij}= 2\langle \Phi_{i}\Phi_{j}\rangle$.
The quantity $Z_{ij}$ is a constant of motion that acts as a type of
$Z_{2}$ gauge field. Closely related identities have recently been used
to solve an anisotropic Heisenberg model on a honeycomb
lattice \cite{kitaev}. 
The  extension of these ideas to a Kondo lattice model, and its possible link to $Z_2$ gauge theories \cite{senthil}
may be of particular interest to future research. 

We should like to thank O. Parcollet for extensive discussions related
to this work. This  work was supported by DOE grant DE-FG02-00ER45790
(PC, DL, WM), and the EPSRC fellowship GR/M70476 (CH).  After
posting this work, we discovered that Shnirman and Makhlin have independently arrived at similar
conclusions\cite{shnirman}.


\end{document}